\documentclass[prd,aps,preprint,tightenlines,nofootinbib,superscriptaddress]{revtex4}
\usepackage{mathrsfs}
\usepackage{amsfonts}
\usepackage{amsmath}
\usepackage{array}
\usepackage{verbatim}
\usepackage{epsfig}
\usepackage{graphicx}
\usepackage{color}

\newcommand{\nn}{\nonumber}

\preprint{
{\vbox {
\hbox{\bf MSUHEP-18-001}
}}}
\vspace*{0.2cm}

\begin{document}
\title{Soft Gluon Resummation in Higgs Boson Plus Two Jet Production at the LHC}

\author{Peng Sun}
\email{pengsun@msu.edu}
\affiliation{Department of Physics and Institute of Theoretical Physics, Nanjing Normal University, Nanjing, Jiangsu, 210023, China }
\affiliation{Department of Physics and Astronomy, Michigan State University,
East Lansing, MI 48824, USA}
\author{C.-P. Yuan}
\email{yuan@pa.msu.edu}
\affiliation{Department of Physics and Astronomy, Michigan State University,
East Lansing, MI 48824, USA}
\author{Feng Yuan}
\email{fyuan@lbl.gov}
\affiliation{Nuclear Science Division, Lawrence Berkeley National
Laboratory, Berkeley, CA 94720, USA}

\begin{abstract}
In this paper, the soft gluon resummation effect in the Higgs boson plus two-jet production
at the LHC is studied. By applying the transverse momentum
dependent factorization formalism, the large logarithms introduced by the small total transverse momentum
of the Higgs boson plus two-jet final state system, are resummed
to all orders in the expansion of the strong interaction coupling at the accuracy of Next-to-Leading Logarithm.
We also compare our result with the prediction of the Monte Carlo event generator Pythia8, and found
noticeable difference in the distributions of the total transverse momentum and the azimuthal angle
correlations of the final state Higgs boson and two-jet system.

\end{abstract}

\maketitle

{\it Introduction.}
After the discovery of Higgs boson at the CERN Large Hadron Colllider (LHC) ~\cite{Aad:2012tfa,Chatrchyan:2012ufa},
determining its properties has become one of the most important tasks for the
high energy physics community.
 It requires a careful comparison between the
experimental measurements of various Higgs boson production and decay channels and
the Standard Model (SM) predictions.
Among these channels, the Higgs boson plus two jet production at the LHC is one of the most important
ones to test the couplings of the Higgs boson~\cite{Aad:2014lwa,Aad:2014tca,Aad:2014eha,Campbell:2006xx,Campbell:2012am,Figy:2003nv,Ellis:2005qe,Dittmaier:2012vm,Arnold:2008rz,Forshaw:2007vb}, which can be expressed as:
\begin{equation}
A(P )+B(\bar P)\to H(P_H)+Jet(P_{J1})+Jet(P_{J2})+X \ ,\label{eq1}
\end{equation}
where $P$ and $\bar P$ represent the incoming hadrons' momenta,
$P_H$ is the momentum of the final state Higgs boson, and the momenta of the final state jets are $P_{J1}$ and $P_{J2}$
with their rapidities $y_{J1}$ and $y_{J2}$, respectively.
In this production, the Higgs boson can be produced via two gluon fusion (GF) or two vector boson fusion (VBF) mechanisms. Being able to separate the GF and VBF production channels would help determining the couplings of Higgs boson, for they are sensitive to the effective coupling of Higgs boson to gluons and to weak gauge bosons, respectively.
To achieve this gaol, we study the kinematic distributions of the Higgs boson and  the two final state jets and their correlations.
For example, the rapidity gap of the two final state jets ($|\Delta y_{JJ}|=|y_{J1}-y_{J2}|$)
in the GF process tends to be smaller than that in the VBF process.
Therefore, requiring a larger value of $|\Delta y_{JJ}|$ would enhance the relative contribution from VBF process
~\cite{Kleiss:1987cj}.

In addition, the differential cross sections of the total transverse momentum ($\vec{q}_\perp=\vec{P}_H^{\perp}+\vec{P}_{J1}^{\perp}+\vec{P}_{J2}^{\perp}$) for the Higgs and
the two final state jets are also sensitive to the production mechanisms.
Such $q_\perp$ distributions are strongly dependent on the soft gluon radiations, especially in the small $q_\perp$ region.
Since the effect of soft gluon radiations is determined by the color structures of  the scattering processes, and
the Higgs boson GF and VBF production mechanisms in this channel have different color structures,
their $q_\perp$ distributions will peak at the different values.
Hence, a precise theoretical prediction on the $q_\perp$ distribution is needed to separate the GF and VBF production processes.
To reliably predict the $q_\perp$ distribution,
soft gluon shower effect must be considered.
The soft gluon shower effect brings the large Sudakov logarithms into
all orders of the perturbative expansion, and then breaks the validity of the perturbative expansion.
Fortunately, this problem can be resolved by performing an all-order transverse momentum dependent (TMD)
resummation calculation based on the TMD factorization theorem~\cite{Collins:1984kg,Ji:2004wu,Collins:2011zzd}, which is
widely used to resum these large logarithms in color
singlet processes~\cite{Collins:1981uk}. For the processes with more
complicated color structures, the TMD resummation was discussed firstly in Ref~\cite{Zhu:2012ts} for colored heavy particle
production processes. For the processes with massless jets in the final states,
the extra soft gluon radiation could be within or outside
the jet cone. Within the jet cone, the radiated gluon can
be treated as a part of the jet, and it
leads to a contribution for the bin of $q_\perp = 0$.
If it is outside the jet cone, it will generate the large Sudakov logarithm,
and it should be resummed.
Its details can be found in the recently developed TMD resummation method~\cite{Sun:2014gfa,Sun:2016kkh,Sun:2016mas}.

In this work, we will apply the TMD resummation method to study
the soft gluon resummation effect on the production of
Higgs boson plus two jets in hadron collision.
In terms of the TMD factorization formalism, the $q_\perp$ differential cross section
is factorized into several individual factors which will be analytically calculated
up to the one-loop order.
As of today, the Monte Carlo (MC) event generators
are the only available tools to predict the soft gluon shower effect for this channel.
Our calculation, for the first time, provides an important test on the validity of the commonly used
MC event generators.

{\it TMD factorization.}
In our calculation, the effective Lagrangian in the heavy top
quark mass limit is used to describe the effective
coupling between Higgs boson and gluons ~\cite{Dawson:1990zj},
 \begin{equation}
{\cal L}_{eff}=-\frac{\alpha_s}{12 \pi v} F^a_{\mu\nu}F^{a\mu\nu}H,
\label{eq:ggh}
\end{equation}
where $v$ is the vacuum expectation value, $H$ the Higgs boson field,
$F^{\mu\nu}$ the gluon field strength tensor, and $a$ the color index.
Our TMD resummation formula can be written as:

\begin{eqnarray}
\frac{d^6\sigma}
{dy_H dy_{J1}dy_{J2} d P_{J1\perp}^2 d P_{J2\perp}^2
d^2\vec{q}_{\perp}}=\sum_{ab}\left[\int\frac{d^2\vec{b}}{(2\pi)^2}
e^{-i\vec{q}_\perp\cdot
\vec{b}}W_{ab\to Hcd}(x_1,x_2,b)+Y_{ab\to Hcd}\right] \ ,\label{resumy}
\end{eqnarray}
where $y_H$, $y_{J1}$ and $y_{J2}$ denote the rapidities of the Higgs boson and the jets, respectively,
$P_{J1\perp}$ and $P_{J2\perp}$ are the jets transverse momentum,
 and $\vec{q}_\perp=\vec{P}_{H\perp}+\vec{P}_{J1\perp}+\vec{P}_{J2\perp}$ is the imbalance
transverse momentum of the Higgs boson and the two final state jets.
The first term ($W$) contains all order resummation effect
and the second term ($Y$) accounts for the difference between
the fixed order result and the so-called asymptotic
result which is given by expanding the resummation result
to the same order in $\alpha_s$ as the fixed order term.
$x_1$ and $x_2$ are the momentum fractions of
the incoming hadrons carried by the incoming partons, with
\begin{eqnarray}
x_{1,2}=\frac{\sqrt{m_H^2+P^2_{H\perp}}e^{\pm y_H}+\sqrt{P^2_{J1\perp}}e^{\pm y_{J1}}+\sqrt{P^2_{J2\perp}}e^{\pm y_{J2}}}{\sqrt{S}} \ .
\end{eqnarray}
We can write the all order resummation
result for $W$ as
\begin{eqnarray}
W_{ab\to Hcd}\left(x_1,x_2,b\right)&=&x_1\,f_a(x_1,\mu_F=b_0/b_*)
x_2\, f_b(x_2,\mu_F=b_0/b_*) e^{-S_{\rm Sud}(Q^2,\hat{\mu})}e^{-\mathcal{F}_{NP}(Q^2,b)} \nonumber\\
&\times& \textmd{Tr}\left[\mathbf{H}_{ab\to Hcd}(\hat{\mu})
\mathrm{exp}[-\int_{b_0/b_*}^{\hat{\mu}}\frac{d
\mu}{\mu}\mathbf{\gamma}_{}^{s\dag}]\mathbf{S}_{ab\to Hcd}(b_0/b_*)
\mathrm{exp}[-\int_{b_0/b_*}^{\hat{\mu}}\frac{d
\mu}{\mu}\mathbf{\gamma}_{}^{s}]\right]\ ,\nonumber\\\label{resum}
\end{eqnarray}
where $s=x_1x_2S$, and $S$ is the hadronic center of mass energy squared,
$b_0=2e^{-\gamma_E}$, with $\gamma_E$ being the Euler constant, $\hat{\mu}$ is the
resummation scale to apply the TMD factorization in the resummation calculation,
as in the Collins 2011 scheme~\cite{Collins:2011zzd}.
$f_{a,b}(x,\mu_F)$ are the parton distribution functions (PDFs) for the incoming
partons $a$ and $b$, and the $\mu_F$ is the evolution scale of the PDFs.
The renormalization scale has been set as the mass of Higgs boson.
$\mathbf{H}_{ab\to Hcd}$ is the hard factor, and it is a matrix based on a set of basis color factors.
By applying the Catani-De Florian-Grazzini(CFG) scheme~\cite{Bozzi:2005wk} and the TMD factorization in the Collins 2011 scheme~\cite{Collins:2011zzd}, we obtain the color singlet component in the hard
factor matrix $H_{ab\to Hcd}^{VBF}$ for the VBF channels, at the next-to-leading order (NLO), as
\begin{eqnarray}
H^{(1)VBF}_{ab\rightarrow Hcd}&=&H^{(0)VBF}_{ab\rightarrow Hcd}\frac{C_F\alpha_s}{2\pi}\left[ -\ln^2 \left( \frac{\hat{\mu}^2}{ \hat{t}_1 }  \right) - \ln^2 \left( \frac{\hat{\mu}^2}{ \hat{t}_2 }  \right)
                                -3 \ln \left( \frac{\hat{\mu}^2}{ \hat{t}_1 }  \right) -3 \ln \left( \frac{\mu^2}{ \hat{t}_2 }  \right)
                                -16  \right.  \nn \\
                              &+&  \frac{1}{2}\ln^2\left(\frac{\hat{\mu}^2}{P^{2}_{J1\perp}}\right) + \frac{3}{2}\ln\left(\frac{\hat{\mu}^2}{R^2P^{2}_{J1\perp}}\right)-\ln\left(R^2\right)\ln\left(\frac{\hat{\mu}^2}{P^{2}_{J1\perp}}\right) +\frac{13}{2}-\frac{2}{3}\pi^2  \nn \\
                              &+&\left.\frac{1}{2}\ln^2\left(\frac{\hat{\mu}^2}{P^{2}_{J2\perp}}\right) + \frac{3}{2}\ln\left(\frac{\hat{\mu}^2}{R^2P^{2}_{J2\perp}}\right)
                                         -\ln\left(R^2\right)\ln\left(\frac{\hat{\mu}^2}{P^{2}_{J2\perp}}\right) +\frac{13}{2}-\frac{2}{3}\pi^2    \right]\ ,
\end{eqnarray}
where $R$ and $P_{J\perp i}$ denote the jet size and transverse momenta of the final state jets, and $H^{(0)}$ is the tree level cross section.
Denoting the initial parton momenta as ($p_a,\;p_b$)
and the final state jet momenta as ($p_{j1},\;p_{j2}$), the above kinematic variables $\hat{t}_1$ and $\hat{t}_2$ can be expressed as $\hat{t}_1= (p_a-p_{j1})^2 $ and $\hat{t}_2= (p_b-p_{j2})^2 $.
Similarly, we have  $\hat{u}_1=(p_a-p_{j2})^2$ and $\hat{u}_1=(p_b-p_{j1})^2$.
In the above hard factor, besides the contribution from the virtual correction at the one-loop order, we have also
included two pieces of contributions from real gluon radiation. The first one is from the jet function,
which describes the gluon radiation within the jet~\cite{Mukherjee:2012uz}.
Here, we follow Ref.~\cite{Mukherjee:2012uz} and apply the dimensional regularization method to integrate the allowed phase space volume of the radiated gluon, and the anti-$k_T$ jet algorithm is adopted.
Another one comes from the $\epsilon$-expansion terms in the soft gluon radiation out
of the jet, which contributes to a finite term when Fourier transformed into
$b$-space with the dimension $D=4-2\epsilon$~\cite{Sun:2014gfa}. Here,
we make the light jet off-shell to regulate the light cone singularity.
As found in Ref.~\cite{Sun:2016kkh}, the different treatment of the jet part in the
 the jet functions and the soft factor leads to
 a finite contribution in the hard factor,
which does not depend on the jet
cone size. Numerically, it is found to be approximately
$\frac{\alpha_s}{2\pi}C_{A}(\frac{\pi^2}{6})$
and
$\frac{\alpha_s}{2\pi}C_{F}(\frac{\pi^2}{6})$
 for gluon and quark jet, respectively.
 We have included these additional contributions
 in the above equation, which were not included in Ref.~\cite{Sun:2016mas}.
For the VBF channel,
in addition to the color singlet contribution, there is also a non-zero
color-octet component in the hard factor matrix, at the $\alpha_s$ order.
However, in the perturbative expansion it does not contribute until next-to-next-leading-order (NNLO).
Hence, we ignore its contribution in this work.

For the GF channel,  we analytically calculate the complete hard factor matrix for the process $H_{gg\to Hgg}^{GF}$, at the NLO,
using the helicity amplitudes given in Ref~\cite{Badger:2009hw}, as suggested in Ref~\cite{Moult:2015aoa}.
The needed color basis for this calculation is identical to that for describing di-jet production in hadron-hadron collision,
as given in Ref~\cite{Sun:2014gfa}.
The soft factor $\mathbf{S}_{ab\rightarrow Hcd}$ is also a matrix in the color space,
and $\gamma^s$ is the associated anomalous dimension for the soft factor,
which can also be obtained
from the result for di-jet production~\cite{Sun:2014gfa} by switching the Mandelstam variables $\hat{t}$ and $\hat{u}$
to $(\hat{t}_1+\hat{t}_2)/2$ and $(\hat{u}_1+\hat{u}_2)/2$, respectively.
The Sudakov form factor ${\cal S}_{\rm Sud}$
resums the leading double logarithms and the sub-leading logarithms, which is
\begin{align}
S_{\rm Sud}(s,\hat{\mu},b_*)=\int^{\hat{\mu}^2}_{b_0^2/b_*^2}\frac{d\mu^2}{\mu^2}
\left[\ln\left(\frac{s}{\mu^2}\right)A+B +D\ln\frac{s}{P_{J\perp1}^2R^2}+
D\ln\frac{s}{P_{J\perp2}^2R^2}\right]\ , \label{su}
\end{align}
where the coefficients $A$, $B$ and $D$ can be expanded
perturbatively in $\alpha_s$.
For GF process, $gg\to Hgg$ process at the NLO,
we have $A=C_A \frac{\alpha_s}{\pi}$, $B=-2C_A\beta_0\frac{\alpha_s}{\pi}$
and $D=C_A\frac{\alpha_s}{2\pi}$. For VBF process, $qq\to Hqq$ process at the NLO,
we have
 $A=C_F \frac{\alpha_s}{\pi}$, $B=-3/2C_F\frac{\alpha_s}{\pi}$ and
$D=C_F\frac{\alpha_s}{2\pi}$.
The coefficients $A$ and $B$ come from the energy evolution effect in the TMD PDFs ~\cite{Ji:2004wu},
so that they only depend on the flavor of the incoming partons and are
independent of the scattering process.
The coefficient $D$ is derived from the soft factor associated with the final state jet.
It quantifies the effect of soft radiation which goes outside the
jet cone, hence it depends on the jet size $R$.
Since our calculation is based on the small cone size approximation,
only the term proportional to $\log(1/R^2)$ is kept in the final expression
of the Sudakov factor of Eq.~(\ref{su}), which describes the $q_\perp$
distribution. The $b$-space variable $b_*=b/\sqrt{1+b^2/b_{\rm{max}}^2}$ with $b_{\rm {max}}=1.5~{\rm GeV}^{-1}$, which make
the lower limit in the Sudakov integrand to be lager than the scale $\Lambda_{QCD}$ and all the pieces in it can
be calculated by the perturbative QCD theory.
Consequently, a non-perturbative factor $e^{-\mathcal{F}_{NP}(Q^2,b)}$
has to be added to model the non-perturbative contribution arising from the large $b$-region.
In this work, we choose the non-perturbative
formalism presented in Ref~\cite{Su:2014wpa}, which however only affects the prediction in extreme small $q_\perp$ region ($q_\perp<1$ GeV).

{\it Numerical Analysis.}
We apply the above resummation formula to compute
the $q_\perp$ distributions of the Higgs boson
production associated with two high energy jets.
In our numeric calculations, we have included the $A^{(2)}$ contribution at the two-loop order~\cite{Catani:2000vq} in the Sudakov form factor,
in addition to the $A^{(1)}$, $B^{(1)}$ and $D^{(1)}$ contributions discussed above.
This is because the coefficient $A^{(2)}$ only depends on the flavor of the incoming
partons, and not on the scattering process.
Besides, we have included a theta-function
$\Theta ( \hat{\mu}-q_\perp )$ in Eq.(3) to limit the range of $q_\perp$
integration.
This results in a similar total cross section predicted from our resummation
calculation as that from the fixed order calculation. Based on the study in Ref~\cite{Sun:2016kkh},
we choose $\hat \mu= P^{lead}_{J\perp}$ or $\hat \mu= P^{sub}_{J\perp}$ in this work
to estimate the theoretical uncertainty, where $P^{lead}_{J\perp}$ and $P^{sub}_{J\perp}$
are the transverse momenta of the final state leading jet and sub-leading jet, respectively. The uncertainty from the choice of $\hat \mu$ will
 decrease after we include higher order corrections in the hard factor.
In addition, we take the mass of the Higgs boson ($m_H$)
to be 125 GeV, and
set the renormalization scale related to the $\alpha_s(\tilde{\mu})$ in the
hard factor to $\tilde{\mu}=m_{H}$ in this study, with the CT14 NNLO PDFs~\cite{Dulat:2015mca}.
Following the experimental analysis~\cite{ATLAS:2016nke},
we require the rapidity of the observed jets to satisfy $|y_J|<4.4$.
We use the anti-$k_t$ algorithm to define the observed jets, and the jet size and the minimal transverse
momentum are set at $R=0.\textrm{4}$ and $P_{J\perp}>30$ GeV.
In our calculation we have applied the narrow jet approximation~\cite{Furman:1981kf}.
We have also constrained the two final state
jets to have a large rapidity separation with $|\Delta y_{JJ}|>2.6$, which is also the region that experimentalists are interested in.

\begin{figure}[tbp]
\includegraphics[width=5.0cm]{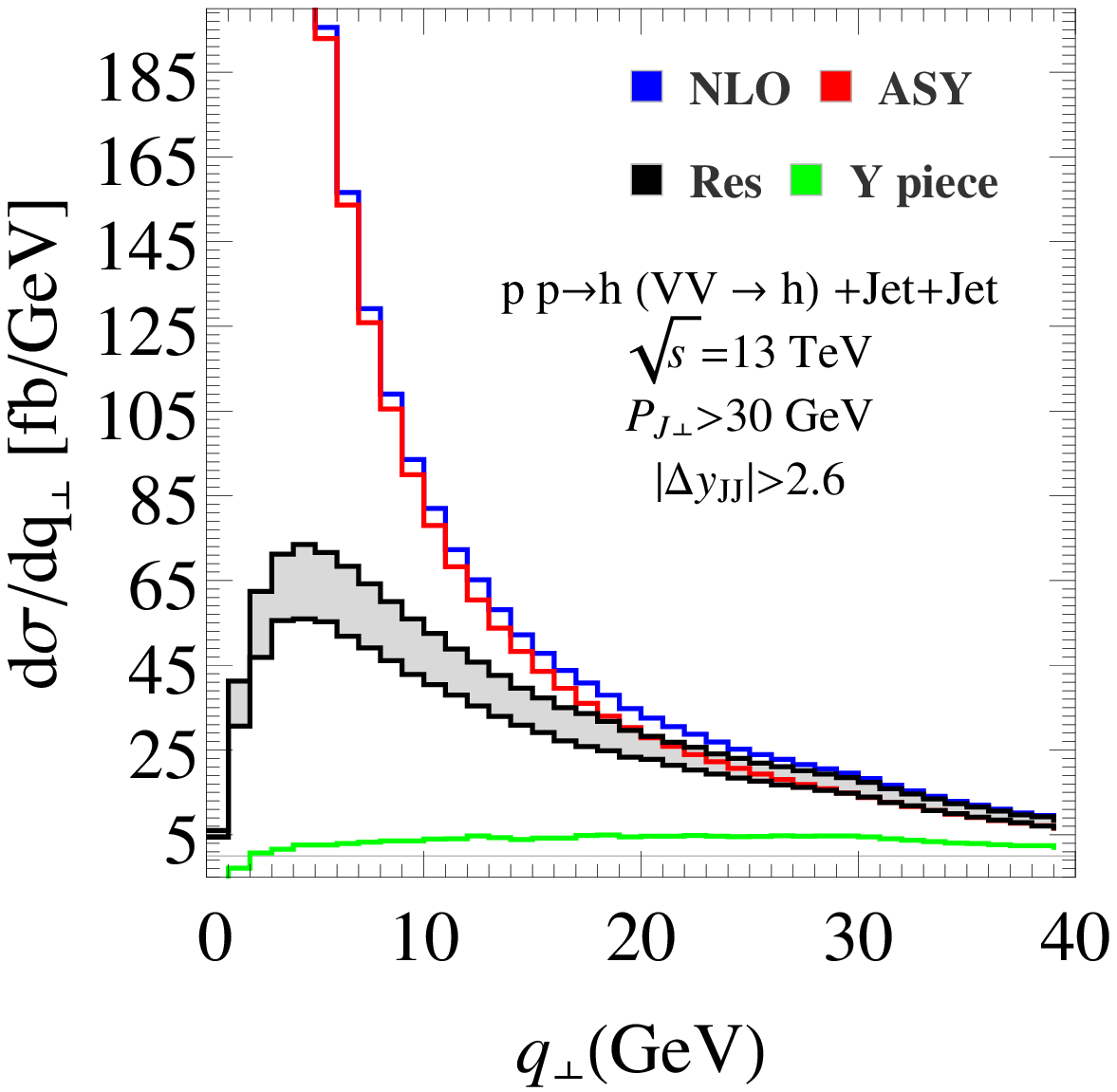}
\includegraphics[width=5.0cm]{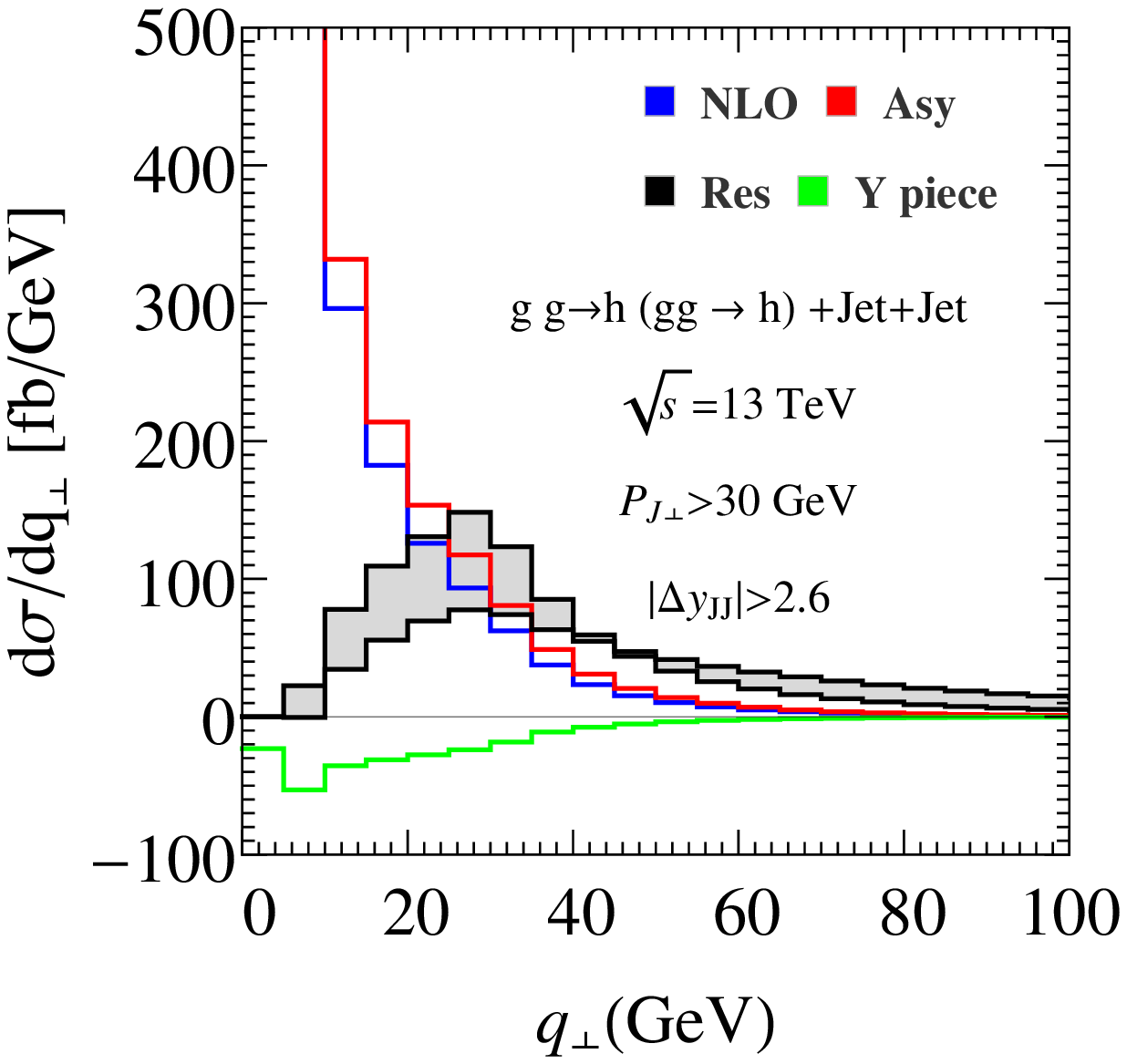}
\includegraphics[width=5.0cm]{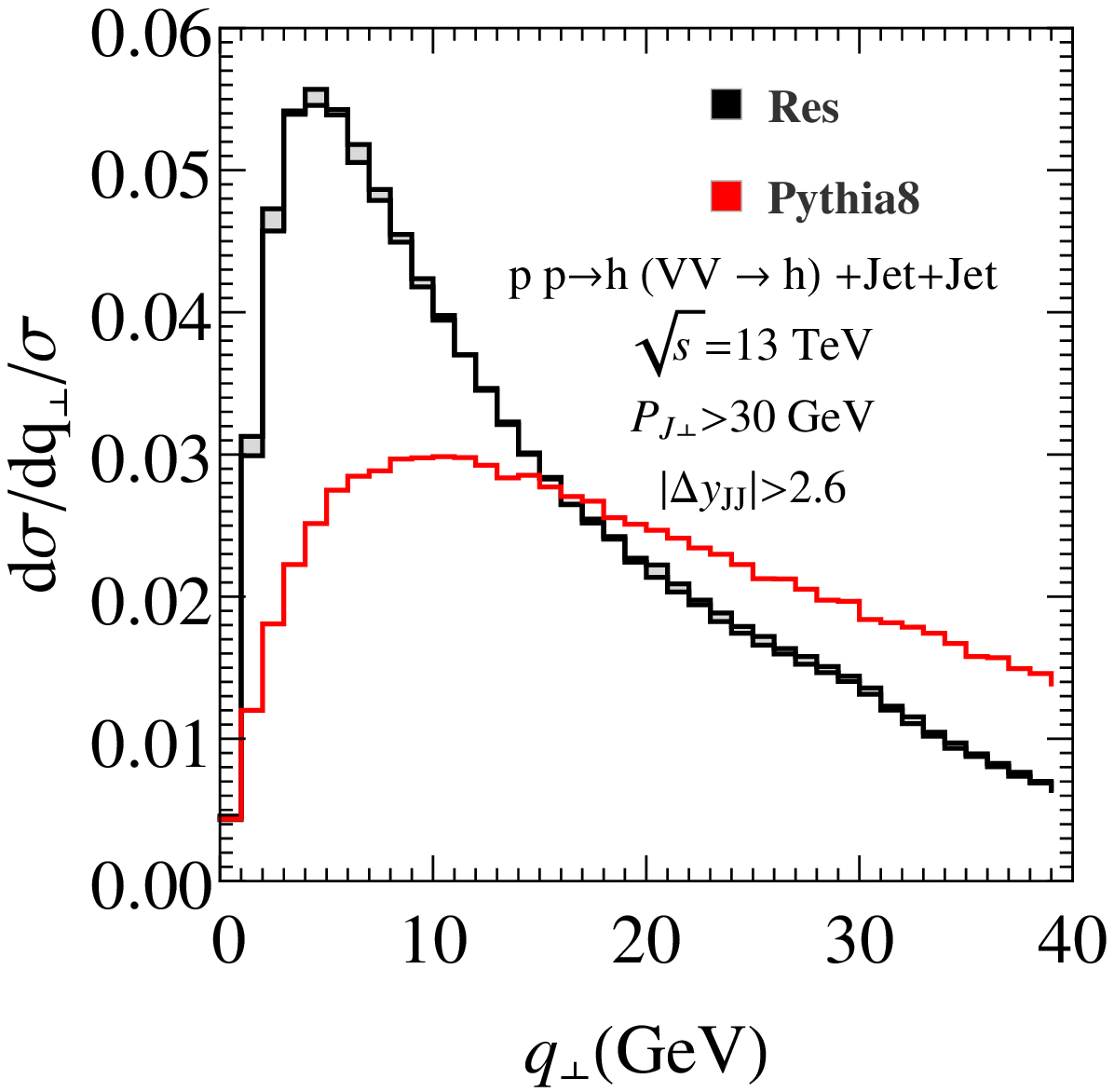}
\includegraphics[width=5.0cm]{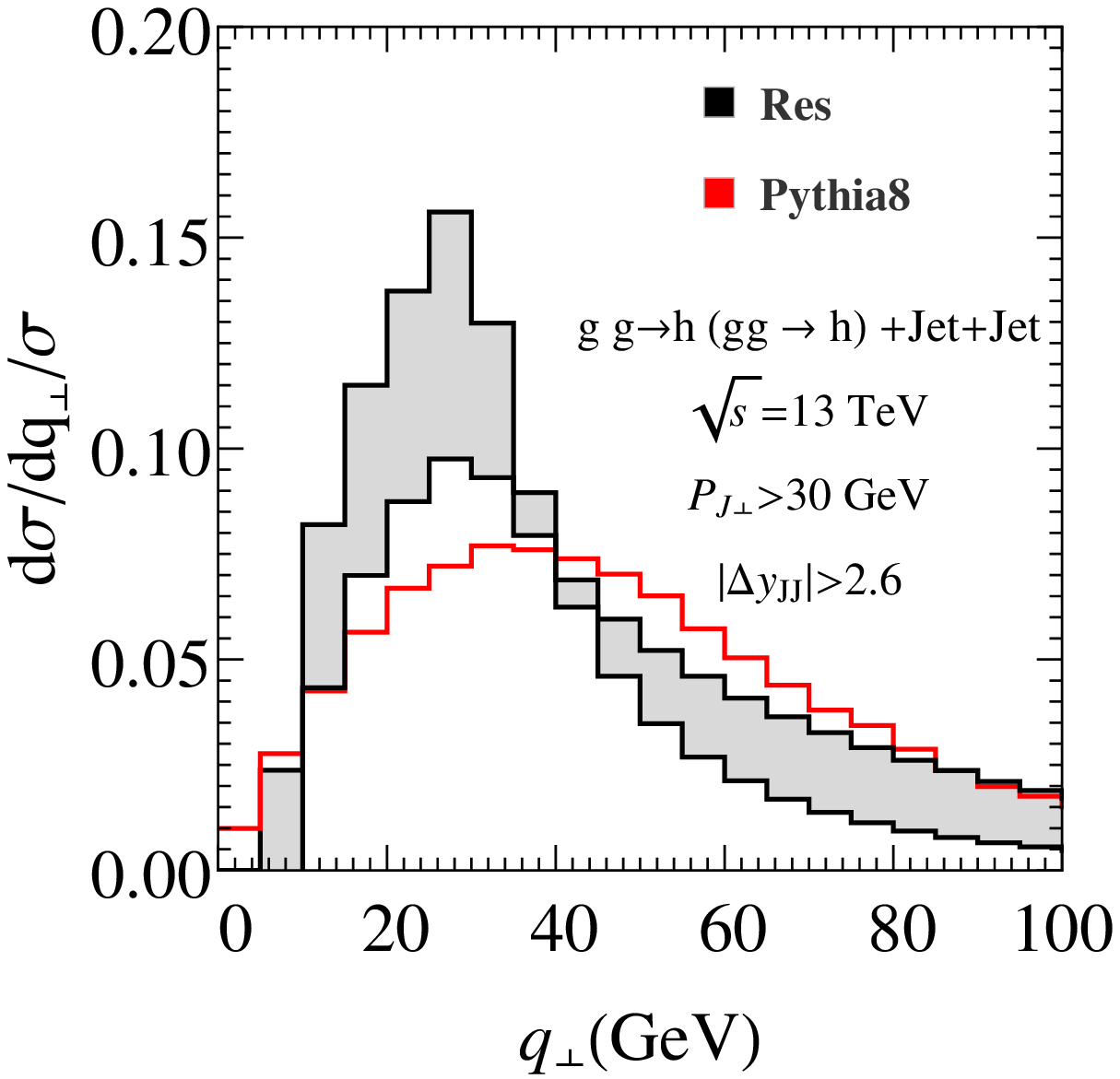}
\includegraphics[width=5.0cm]{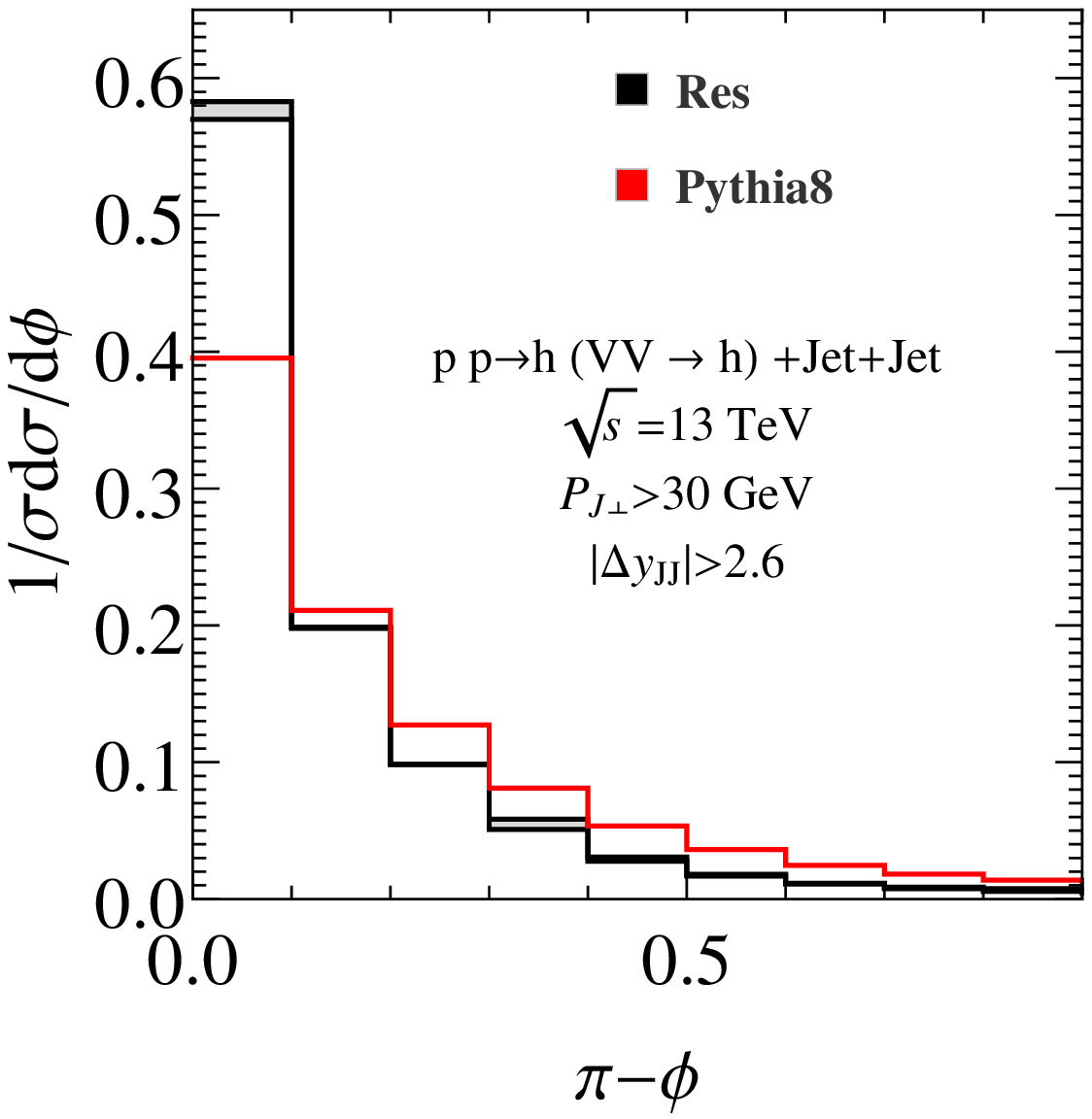}
\caption{
The differential cross sections of Higgs boson plus two jet production
at the LHC as functions of $q_\perp$ and azimuthal angle $\phi$ between
Higgs boson and the final state two-jet system.
In these plots, the $\alpha_s$ order $Y$ pieces are included in the resummation curves.
The predictions from Pythia8 are based on the tree level scattering amplitudes with parton showers.
The uncertainty of our resummation calculation is estimated by varying the resummation scale $\hat \mu$ from $P^{lead}_{J\perp}$ to $P^{sub}_{J\perp}$. }
\label{asymptotic}
\end{figure}

Finally, we compare in Fig.\,1 the predictions from our resummation calculation,
the fixed
order calculation, and the MC event generator Pythia8.
A noticeable difference is found in the shape of the $q_\perp$ distribution, predicted from our resummation calculation, for the VBF and GF production processes.
The peak position for the VBF process is around 5 GeV, while for the GF process, it is around 30 GeV.
This is because the sub-leading logarithm in the VBF production process can become
large, as analysed in Ref~\cite{Sun:2016mas},
which can then push the peak position of the $q_\perp$ distribution
to a much smaller value than that in the GF production process.
Hence, the GF contribution can be largely suppressed, as compared to the VBF contribution,  by requiring $q_\perp$ to be small in the Higgs boson plus two jet events produced at the LHC.
To further compare the Pythia8 predictions against ours,
besides the $q_\perp$ differential cross sections in the first two plots of Fig.\,1,
we also show their normalized distributions in the third and fourth plots.
It is evident that Pythia8 predicts a flatter shape than ours, and another significant
disagreement lies in the peak position of the $q_\perp$ distribution from the
VBF production process.
Pythia8 predicts a peak in $q_\perp$ around $10$ GeV, while ours is at about $5$ GeV.

In the fifth plot of Fig.~1, we also show the distribution of the azimuthal angle between
the Higgs boson and the final state jet pair. Such differential cross section is also sensitive to the
soft gluon radiation and the Higgs boson production mechanism. Requiring a large separation
in this azimuthal angle could largely suppress the GF contribution, but not the VBF contribution.
The experimentalists at the LHC have already applied this technique to
enhance the fraction of VBF contribution in their
data, after imposing some proper kinematic cuts ~\cite{ATLAS:2016nke},
in order to measure the coupling
of Higgs boson to weak gauge bosons.
To this aim, a precise theoretical evaluation of the kinematic acceptance after imposing the
kinematic cuts is needed.
In Ref.~\cite{ATLAS:2016nke}, the ATLAS Collaboration required the azimuthal angle
separation ($\phi$) between the Higgs boson and the di-jet system to be
$\phi>2.6$, and compared the measured fiducial cross section with the Pythia8~prediction.
Below, we shall compare the predicted kinematic acceptance
from Pythia8 to our resummation calculation.
As shown in Table I, the predicted kinematic acceptance
with $\phi>2.84$ is larger by about 14\% in
our resummation calculation than in Pythia8.
For $\phi>2.6$, they differ by about 8\%, and
our resummation calculation results in a larger
total fiducial cross section.
This implies a larger value in the coupling of Higgs boson to weak gauge bosons by about 4\%.
At the High-Luminosity LHC, with an integrated luminosity of up to
3000 fb$^{-1}$, the expected precision on the measurement of the production cross section of
the SM-like Higgs boson via VBF mechanism is around 10\%~\cite{CMS:2017cwx}.
Hence, the difference found in our resummation and Pythia8 calculations of the fiducial cross
sections could become important. Further comparisons on various event shapes between
the experimental data and our resummation predictions could also be carried out in order
to test the Standard Model and to search for New Physics.

 \begin{table}
\caption{ The predicted kinematic acceptances for the azimuthal angle cut-off in the Higgs boson plus two jet production at the LHC.}
\label{tbl:cut}
\begin{tabular}{c|c|c|c|c|c}
\hline
cut-off($\pi-\phi $)  &$<$0.2&$<$0.3&$<$0.4& $<$0.5 &$<$0.54 \\
\hline
Res VBF            &78.15$\sim$76.77\%   &88.00$\sim$86.59\%    &93.09$\sim$92.43\%    &95.85$\sim$95.49\%    &  96.53$\sim$96.50\%\\
\hline
Pythia8 VBF        &60.64\%   &73.35\%    &81.45\%    &86.80\%    &  88.44\% \\
\hline
\end{tabular}
\end{table}

{\it Summary.}
In summary, we have applied the TMD resummation theorem to
study the production of the Higgs boson associated with
two inclusive jets at the LHC. Based on the TMD factorization formalism,
all the factors are calculated up to the NLO. This is the first time
in the literature the effect from multiple soft gluon radiation is studied for this production channel of the Higgs boson
at the accuracy of Next-to-Leading Logarithm.
Our work also provides a
framework for applying the TMD resummation calculation to other $2 \to 3$ scattering processes.
We find large difference between the Pythia8 and our predictions in the
distributions of the total transverse momentum ($q_\perp$) and the azimuthal angle
($\phi$) correlations of the final state Higgs boson and two-jet system,
 after imposing the kinematic cuts
used in the LHC data analysis.

\begin{acknowledgments}
We would like to thank Joshua Isaacson, Bin Yan and Kirtimaan Mohan for helpful discussion.
This work is partially supported by the U.S. Department of Energy,
Office of Science, Office of Nuclear Physics, under contract number
DE-AC02-05CH11231, and by the U.S. National
Science Foundation under Grant No. PHY-1417326.
C.-P. Yuan is also grateful for the support from
the Wu-Ki Tung endowed chair in particle physics.
\end{acknowledgments}


\begin{thebibliography}{99}

\bibitem{Aad:2012tfa}
  G.~Aad {\it et al.}  [ATLAS Collaboration],
  Phys.\ Lett.\ B {\bf 716}, 1 (2012).

\bibitem{Chatrchyan:2012ufa}
  S.~Chatrchyan {\it et al.}  [CMS Collaboration],
  Phys.\ Lett.\ B {\bf 716}, 30 (2012).

\bibitem{Aad:2014lwa}
  G.~Aad {\it et al.}  [ATLAS Collaboration],
  arXiv:1407.4222 [hep-ex].

\bibitem{Aad:2014tca}
  G.~Aad {\it et al.}  [ ATLAS Collaboration],
  arXiv:1408.3226 [hep-ex].

\bibitem{Aad:2014eha}
  G.~Aad {\it et al.}  [ ATLAS Collaboration],
  arXiv:1408.7084 [hep-ex].


\bibitem{Campbell:2006xx}
  J.~M.~Campbell, R.~K.~Ellis and G.~Zanderighi,
  JHEP {\bf 0610}, 028 (2006)
  [hep-ph/0608194].

\bibitem{Campbell:2012am}
  J.~M.~Campbell, R.~K.~Ellis, R.~Frederix, P.~Nason, C.~Oleari and C.~Williams,
  JHEP {\bf 1207}, 092 (2012)
  [arXiv:1202.5475 [hep-ph]].


\bibitem{Figy:2003nv}
  T.~Figy, C.~Oleari and D.~Zeppenfeld,
  Phys.\ Rev.\ D {\bf 68}, 073005 (2003)
  [hep-ph/0306109].

\bibitem{Ellis:2005qe}
  see, for example, R.~K.~Ellis, W.~T.~Giele and G.~Zanderighi,
  Phys.\ Rev.\ D {\bf 72}, 054018 (2005)
  [Erratum-ibid.\ D {\bf 74}, 079902 (2006)]
  [hep-ph/0506196].





\bibitem{Dittmaier:2012vm}
  S.~Dittmaier, S.~Dittmaier, C.~Mariotti, G.~Passarino, R.~Tanaka, S.~Alekhin, J.~Alwall and E.~A.~Bagnaschi {\it et al.},
  arXiv:1201.3084 [hep-ph];
  S.~Heinemeyer {\it et al.}  [LHC Higgs Cross Section Working Group Collaboration],
  arXiv:1307.1347 [hep-ph].

\bibitem{Arnold:2008rz}
  K.~Arnold {\it et al.},
  Comput.\ Phys.\ Commun.\  {\bf 180}, 1661 (2009)
  doi:10.1016/j.cpc.2009.03.006
  [arXiv:0811.4559 [hep-ph]].

\bibitem{Forshaw:2007vb}
  J.~R.~Forshaw and M.~Sjodahl,
  JHEP {\bf 0709}, 119 (2007)
  doi:10.1088/1126-6708/2007/09/119
  [arXiv:0705.1504 [hep-ph]].

\bibitem{Kleiss:1987cj}
  R.~Kleiss and W.~J.~Stirling,
  Phys.\ Lett.\ B {\bf 200}, 193 (1988).
  doi:10.1016/0370-2693(88)91135-5


  \bibitem{Collins:1984kg}
  J.~C.~Collins, D.~E.~Soper and G.~F.~Sterman,
  Nucl.\ Phys.\ B {\bf 250}, 199 (1985).
\bibitem{Ji:2004wu}
  X.~Ji, J.~P.~Ma and F.~Yuan,
  Phys.\ Rev.\ D {\bf 71}, 034005 (2005);
  JHEP {\bf 0507}, 020 (2005).

\bibitem{Collins:2011zzd}
  J.~Collins,
  ``Foundations of perturbative QCD,''
  (Cambridge monographs on particle physics, nuclear physics and cosmology. 32).


\bibitem{Collins:1981uk}
  J.~C.~Collins and D.~E.~Soper,
  Nucl.\ Phys.\ B {\bf 193}, 381 (1981)
  Erratum: [Nucl.\ Phys.\ B {\bf 213}, 545 (1983)].
  J.~C.~Collins and D.~E.~Soper,
  Nucl.\ Phys.\ B {\bf 197}, 446 (1982).

\bibitem{Zhu:2012ts}
  H.~X.~Zhu, C.~S.~Li, H.~T.~Li, D.~Y.~Shao and L.~L.~Yang,
  Phys.\ Rev.\ Lett.\  {\bf 110}, no. 8, 082001 (2013)
  H.~T.~Li, C.~S.~Li, D.~Y.~Shao, L.~L.~Yang and H.~X.~Zhu,
  Phys.\ Rev.\ D {\bf 88}, 074004 (2013)

\bibitem{Sun:2014gfa}
  P.~Sun, C.-P.~Yuan and F.~Yuan,
  Phys.\ Rev.\ Lett.\  {\bf 113}, no. 23, 232001 (2014).
  Phys.\ Rev.\ D {\bf 92}, no. 9, 094007 (2015).


\bibitem{Sun:2016kkh}
  P.~Sun, J.~Isaacson, C.-P.~Yuan and F.~Yuan,
  Phys.\ Lett.\ B {\bf 769}, 57 (2017)
  doi:10.1016/j.physletb.2017.02.037
  [arXiv:1602.08133 [hep-ph]].



\bibitem{Sun:2016mas}
  P.~Sun, C.-P.~Yuan and F.~Yuan,
  Phys.\ Lett.\ B {\bf 762}, 47 (2016)




\bibitem{Dawson:1990zj}
  S.~Dawson,
  Nucl.\ Phys.\ B {\bf 359}, 283 (1991).


\bibitem{Bozzi:2005wk}
  G.~Bozzi, S.~Catani, D.~de Florian and M.~Grazzini,
  Nucl.\ Phys.\ B {\bf 737}, 73 (2006)


\bibitem{Mukherjee:2012uz}
  A.~Mukherjee and W.~Vogelsang,
  Phys.\ Rev.\ D {\bf 86}, 094009 (2012)


\bibitem{Badger:2009hw}
  S.~Badger, E.~W.~Nigel Glover, P.~Mastrolia and C.~Williams,
  JHEP {\bf 1001}, 036 (2010)

\bibitem{Moult:2015aoa}
  I.~Moult, I.~W.~Stewart, F.~J.~Tackmann and W.~J.~Waalewijn,
  Phys.\ Rev.\ D {\bf 93}, no. 9, 094003 (2016)

\bibitem{Su:2014wpa}
  P.~Sun, J.~Isaacson, C.-P.~Yuan and F.~Yuan,
  arXiv:1406.3073 [hep-ph].

\bibitem{Catani:2000vq}
  S.~Catani, D.~de Florian and M.~Grazzini,
  Nucl.\ Phys.\ B {\bf 596}, 299 (2001).
  S.~Catani, L.~Cieri, D.~de Florian, G.~Ferrera and M.~Grazzini,
  Nucl.\ Phys.\ B {\bf 881}, 414 (2014).


\bibitem{Dulat:2015mca}
  S.~Dulat {\it et al.},
  Phys.\ Rev.\ D {\bf 93}, no. 3, 033006 (2016)
  doi:10.1103/PhysRevD.93.033006
  [arXiv:1506.07443 [hep-ph]].

\bibitem{ATLAS:2016nke}
  The ATLAS collaboration [ATLAS Collaboration],
  ATLAS-CONF-2016-067.









\bibitem{Furman:1981kf}
  M.~Furman,
  Nucl.\ Phys.\ B {\bf 197}, 413 (1982);
  F.~Aversa, P.~Chiappetta, M.~Greco and J.~P.~Guillet,
  Nucl.\ Phys.\ B {\bf 327}, 105 (1989);
  Z.\ Phys.\ C {\bf 46}, 253 (1990);
  D.~de Florian and W.~Vogelsang,
  Phys.\ Rev.\ D {\bf 76}, 074031 (2007).



\bibitem{CMS:2017cwx}
  CMS Collaboration [CMS Collaboration],
  CMS-PAS-FTR-16-002.




\end{thebibliography}
\end{document}